# Evaluating Software Contribution Quality: Time-to-Modification Theory

**Vincil Bishop III**[*]
Department of Systems Engineering
Colorado State University
Vincil.Bishop@colostate.edu

**Steven Simske**
Department of Systems Engineering
Colorado State University
Steve.Simske@colostate.edu

## Abstract

The durability and quality of software contributions are critical factors in the long-term maintainability of a codebase. This paper introduces the Time to Modification (TTM) Theory, a novel approach for quantifying code quality by measuring the time interval between a code segment's introduction and its first modification. TTM serves as a proxy for code durability, with longer intervals suggesting higher-quality, more stable contributions. This work builds on previous research, including the "Time-Delta Method for Measuring Software Development Contribution Rates" dissertation, from which it heavily borrows concepts and methodologies. By leveraging version control systems such as Git, TTM provides granular insights into the temporal stability of code at various levels—ranging from individual lines to entire repositories. TTM Theory contributes to the software engineering field by offering a dynamic metric that captures the evolution of a codebase over time, complementing traditional metrics like code churn and cyclomatic complexity. This metric is particularly useful for predicting maintenance needs, optimizing developer performance assessments, and improving the sustainability of software systems. Integrating TTM into continuous integration pipelines enables real-time monitoring of code stability, helping teams identify areas of instability and reduce technical debt.

## 1 Introduction

The quality and longevity of software contributions play a crucial role in the overall sustainability of a codebase. One approach to quantify the durability of these contributions is through measuring how long it takes for a particular segment of code, or Git "hunk," to be overwritten after its initial introduction. This metric, referred to as Time to Modification (TTM), could serve as a temporal durability indicator, shedding light on the lasting impact of individual developers' contributions.

Traditional version control systems, such as Git, do not offer a built-in mechanism to directly calculate TTM. However, Git's "Blame" feature provides an indirect method by tracing the history of changes to specific lines of code. By leveraging this tool, it is possible to track how long each hunk remains unmodified, thus allowing us to derive a measure of its durability over time.

The central hypothesis behind this approach is that code that remains unmodified for longer periods is generally of higher quality and exhibits greater durability. This idea builds on the assumption that stable, well-written code requires fewer modifications, and therefore, the longer a piece of code remains unchanged, the more durable and effective it is likely to be. Code that has been shared or repurposed in other applications would exemplify this hypothesis as it may require few if any modifications, and by the fact that it was chosen for reuse by default exhibits a level of quality, if not high quality and durability.

Given the granular nature of Git changes, the data produced by this analysis can be extensive, offering a wealth of insights at various levels—from individual contributions to the durability trends of entire developer cohorts.

---

[*] Work does not relate to position at Amazon.

The conceptual model of a durability graph thus emerges as a tool for visualizing and understanding these trends, enabling a more precise evaluation of code quality and developer performance in a temporal context.

## 2  Background and Related Work

A significant portion of the background information and theoretical foundations discussed in this work have been adapted from Bishop's doctoral dissertation titled "Time-Delta Method for Measuring Software Development Contribution Rates." [1] This reuse of concepts and methodologies reflects a continuation of the research conducted in the dissertation and serves as a critical foundation for the development of the Time-to-Modification Theory presented here.

### 2.1  Code Quality in General

Measuring general code encompasses various methodologies, metrics, and tools. The most accepted methods for assessing code quality can be broadly categorized into manual analysis techniques, automated tools, and metrics-based evaluations. Each method has its strengths and weaknesses, and often, a combination of these approaches is employed to achieve a comprehensive understanding of code quality.

Manual analysis techniques, such as code inspections and peer reviews, are foundational practices in software development. These methods involve human evaluators examining the code for potential defects, adherence to coding standards, and overall maintainability. Plösch et al. highlight the effectiveness of expert-based evaluations in identifying internal software quality issues, emphasizing that manual analysis can uncover problems that automated tools might miss [2]. Similarly, Hadyan et al. discuss the importance of code reviews in identifying defects before code integration, although they note that the labor-intensive nature of these reviews can hinder their widespread adoption [3]. Furthermore, Runeson et al. advocate for code inspections as a reliable method to enhance software quality, reinforcing the value of human judgment in the evaluation process [2].

Automated tools have gained prominence in recent years due to their ability to provide consistent and objective assessments of code quality. Tools that implement static code analysis can automatically evaluate various quality metrics, such as cyclomatic complexity, lines of code, and code smells. Zaw et al. identify several key metrics, including the number of classes (NOC), number of methods (NOM), and cyclomatic complexity (CC), which are essential for assessing code quality object oriented languages [4]. These metrics can be integrated into continuous integration pipelines to provide real-time feedback to developers, facilitating a proactive approach to code quality management. Kothapalli et al. introduce the Code Quality Monitoring Method (CQMM), which systematically monitors code quality throughout the development lifecycle, utilizing static analysis tools to ensure adherence to quality standards [5].

Metrics-based evaluations are critical for quantifying code quality and providing actionable insights. Various quality models, such as ISO/IEC 9126, offer frameworks for evaluating software quality characteristics, including functionality, reliability, usability, efficiency, maintainability, and portability [6]. Kapse et al. emphasize the importance of developing precise quality measurement techniques that can yield higher-quality code, suggesting that metrics should be tailored to the specific context of the software being developed [7]. The use of ownership metrics has been explored by Foucault et al., who found a correlation between code ownership and software quality, indicating that understanding the dynamics of code ownership can enhance quality assessments [8].

The integration of automated tools with manual analysis techniques can lead to a more robust evaluation of code quality. For instance, automated code analyzers can provide initial assessments, which can then be supplemented by manual reviews to capture nuances that automated tools may overlook. This hybrid approach is supported by findings from Butler et al., who explored the influence of identifier names on code quality, demonstrating that both automated metrics and human evaluations are necessary for a comprehensive assessment [9]. Additionally, the use of machine learning techniques to analyze code quality has emerged as a promising avenue for future research, as highlighted by Guo et al., who emphasize the need for quantitative evaluations of quality models [10].

### 2.2  Temporal Stability

Measuring the temporal stability of software code is a critical aspect of software engineering, particularly as systems evolve over time. Various approaches have been developed to assess this stability, focusing on different metrics and methodologies that reflect changes in code quality, maintainability, and defect density. This synthesis will explore several prominent approaches, supported by empirical evidence from the literature.

One of the foundational approaches to measuring temporal stability is through the analysis of code churn, which quantifies the frequency and extent of changes made to the codebase over time. Nagappan & Ball [11] introduced relative code churn measures as predictors of system defect density, suggesting that higher churn rates correlate with increased defect rates. This relationship underscores the importance of monitoring code changes as a means of assessing software stability. Similarly, Meneely & Williams [12] highlighted the "code churn" effect, indicating that files with significant changes are more likely to harbor faults and vulnerabilities. These findings suggest that tracking code churn can provide valuable insights into the stability and reliability of software systems.

In addition to code churn, maintainability metrics play a crucial role in assessing temporal stability. Hota et al. Hota et al. [13] emphasized the importance of reliability analysis during the software development life cycle (SDLC), proposing models that utilize source code metrics to predict aging-related bugs. Automated analysis techniques, as discussed by Chen and Shoga [14], involve quantifying maintainability through static code analysis, which includes metrics such as the Maintainability Index and technical debt. These metrics can serve as indicators of how well a codebase can adapt to changes over time, thus reflecting its temporal stability.

Refactoring is another significant approach to improving and measuring the temporal stability of software code. Keenan et al. [15] investigated the impact of refactoring on software evolution, revealing that refactoring activities can reduce entropy within a codebase, thereby enhancing maintainability and stability. This aligns with findings from Szoke et al. [16], who noted that continuous reengineering and refactoring practices are essential for maintaining code quality over time. By systematically addressing code smells and complexity through refactoring, developers can mitigate the risks associated with software decay and improve the overall stability of the code.

The use of automated tools for assessing code quality has gained traction in recent years. Welker et al. [17] developed an automated source code maintainability index that quantifies maintainability through a minimal set of easily calculated metrics. This approach allows for continuous monitoring of code quality, facilitating timely interventions when stability issues arise. Tools like SonarQube[1], as mentioned by Haderer et al.[18], enable developers to track various quality metrics, including technical debt and code smells, providing a comprehensive view of the code's stability over time.

In the context of cybersecurity, these automated tools offer additional benefits. If a vulnerability in widely distributed code is identified, the continuous monitoring capabilities of these tools can expedite the detection of the vulnerable segments across the codebase. By providing a unified and consistent view of the quality and security status of the code, tools like SonarQube can assist in rapidly addressing security vulnerabilities. In distributed systems, this allows teams to apply patches or refactor vulnerable sections more effectively and ensure that security fixes are propagated consistently across the entire distribution, thereby minimizing the risk of exploitation.

The role of testing in maintaining software stability cannot be overstated. Thahseen [19] found that extensive testing reduces code complexity and defects, thereby enhancing maintainability and stability. Regression testing, in particular, is crucial for ensuring that changes do not introduce new defects. Vidács et al. [20] emphasized the importance of effective regression test suites in maintaining software quality throughout continuous changes, suggesting that the quality of testing directly impacts the stability of the codebase.

The concept of technical debt is integral to understanding temporal stability. Saklani [21] highlighted the significance of predicting various properties of software quality, including technical debt, which can accumulate over time and affect maintainability. By employing metrics that assess technical debt, developers can gain insights into the long-term stability of their code and make informed decisions about necessary refactoring or redesign efforts.

In addition to these quantitative approaches, qualitative assessments also play a role in measuring temporal stability. The analysis of code comments and documentation, as explored by Jiang and Hassan [22], can provide insights into the maintainability of a codebase. A well-documented codebase is often more stable, as it facilitates understanding and modifications by future developers. This qualitative aspect complements the quantitative metrics, offering a holistic view of software stability.

## 3    Time to Modification Theory

Time-to-Modification (TTM) Theory proposes that the temporal stability of a software contribution, defined as the length of time between the initial introduction of a code segment (or "hunk"[2]) and its first modification, serves as a proxy for the durability and quality of the contribution.

According to this theory:

- **Durability as Temporal Stability:** Code segments that remain unchanged for longer periods after their introduction demonstrate greater durability. Durability is a reflection of the robustness, correctness, and maintainability of the contribution. Code that is modified sooner after its introduction is less durable, possibly reflecting lower quality or the need for improvement.
- **Time-to-Modification as a Metric:** The TTM metric offers a measurable way to assess code durability across multiple contexts, including individual contributors, teams, repositories, or projects. It captures the lifespan of code in a version-controlled environment before it undergoes its first revision.
- **Higher Durability Implies Higher Quality:** On average, contributions with higher TTM values are hypothesized to be of higher quality, as they require fewer immediate modifications or fixes. Code that remains stable over time is likely to be better designed, easier to maintain, and more resilient to changes in the system or requirements.
- **Granularity and Scale:** The theory accounts for the highly granular nature of version control systems like Git, enabling a large volume of TTM data points. This allows for robust analysis of code durability patterns at various levels of granularity—from individual code snippets to entire repositories.
- **TTM as a Predictor of Maintenance Needs:** Contributions with low TTM values (i.e., quickly modified) may signal areas of the codebase that are less stable and may require more ongoing maintenance, while high TTM values suggest stable, long-lasting contributions.

---

[1] https://www.sonarsource.com/products/sonarqube/
[2] https://libgit2.org/libgit2/#HEAD/type/git_diff_hunk

This theory provides a structured framework to evaluate the longevity and quality of software contributions, using TTM as a key metric. It offers a quantitative basis for assessing and comparing the durability of code contributions in various software development contexts, ultimately guiding practices that improve software sustainability and quality.

## 3.1 Procedure for Measuring Time to Modification (TTM)

The Time to Modification (TTM) of software contributions can be quantified by analyzing the modification patterns of code segments (hunks) across sequential Git commits. This method tracks the lifespan of a code hunk by determining the time interval between its introduction in an initial commit and the first subsequent modification.

The process begins by collecting commits in chronological order, from oldest to newest, along with associated file patches and their respective hunks. Each hunk is assigned a unique identifier, which is composed of key information including the commit SHA, file path, starting line number, and ending line number. The commit timestamp is also recorded to allow for temporal calculations.

Once a hunk is identified, a Git Blame operation is performed to trace the history of the lines encompassed by the hunk. The blame operation is scoped to the specific commit, file, and line range associated with the hunk. The Blame process yields references to prior commits that modified the same lines, or no result if the hunk represents entirely new lines.

For each BlameHunk returned, the system checks the hunk index to determine whether a modification time has already been recorded. If not, the commit timestamp of the first modifying commit is recorded as the first modification time for the hunk. The Time to Modification is calculated as the difference between the timestamp of the hunk's introduction and the timestamp of its first modification, with the time delta expressed in seconds.

Since the process iterates through commits in chronological order, the first time a modification is identified for a given hunk will correspond to its TTM. Subsequent modifications are disregarded for the purposes of this measurement, as only the initial modification is required to determine durability.

## 3.2 Mean Time to Modification (MTTM)

The Mean Time to Modification (MTTM) is a statistical measure used to quantify the average durability of code contributions in a software repository. It builds on the previously defined Time to Modification (TTM) metric, which captures the lifespan of individual code segments (hunks) from their introduction until their first modification. By calculating the mean of TTM values across all code contributions, the MTTM provides insight into the general stability and quality of the codebase.

### 3.2.1 MTTM Mathematical Foundations

The MTTM reflects the collective durability of software contributions by averaging the time intervals between the initial introduction of code hunks and their first subsequent modification. A higher MTTM suggests that the codebase is relatively stable, with contributions remaining unchanged for longer periods before requiring modification. This could be indicative of well-designed and robust code. Conversely, a lower MTTM indicates that code modifications are occurring more frequently, suggesting areas of the codebase that may require closer attention for potential refactoring, bug fixing, or enhancement.

The theory underlying MTTM is grounded in the assumption that contributions which remain stable for extended periods are of higher quality. Such code typically requires fewer revisions, implying that it may have been designed and implemented effectively from the outset. The metric is particularly useful in contexts where software maintainability and long-term stability are prioritized, providing an aggregate view of how well the codebase is evolving over time.

### 3.2.2 Calculation of MTTM

The MTTM is computed by taking the mean of all individual TTM values across a set of contributions. The calculation proceeds as follows:

Collect TTM Values: For each code hunk in the repository, compute the TTM as the time interval between the commit that introduces the hunk and the commit that first modifies it. This is expressed in a suitable time unit, such as seconds, minutes, or days.

$$TTFM_i = t_{modified} - t_{introduced}$$

Where:

- $TTFM_i$ is the Time to Modification for hunk.
- $t_{modified}$ is the timestamp of the first modifying commit.
- $t_{introduced}$ is the timestamp of the commit that introduces the hunk.

Aggregate TTM Values: Once TTM values have been calculated for all relevant code hunks, aggregate these values.

Calculate the Mean: The MTTM is the arithmetic mean of all collected TTM values. This can be expressed as:

$$MTTFM = \frac{1}{n}\sum_{i=1}^{n} TTFM_i$$

Where:

- $MTTFM$ is the mean time to first modification.
- $TTFM_i$ is the Time to Modification for the $i$-th hunk.
- $n$ is the total number of hunks analyzed.

Interpret the Results: The resulting MTTM value provides a summary statistic for the repository or a subset of the repository. A higher value suggests greater code durability, while a lower value indicates more frequent modifications and, potentially, less stable contributions.

### 3.2.3 Practical Considerations

- **Granularity of Data:** The MTTM is highly dependent on the granularity of the TTM calculations. Analyzing code hunks at the level of individual lines or small patches will yield finer-grained results compared to larger hunks.
- **Temporal Resolution:** The time unit used for TTM (e.g., seconds, minutes, days) should be selected based on the scale of the development process. For high-frequency, agile projects, a smaller time unit may be more appropriate, while larger time units may be better suited for long-term projects with infrequent modifications.
- **Outliers:** Care should be taken when interpreting MTTM values, as outliers (such as extremely durable code or rapidly modified code) can skew the mean. In such cases, it may be useful to calculate additional statistics, such as the median or standard deviation, to gain a more comprehensive understanding of the data.

# 4 Practical Implications

## 4.1 Processing Challenges

The computational process of measuring Time to Modification (TTM) faces significant scalability challenges, particularly due to the need to index and reference each individual hunk that has already been processed. The sheer volume of data involved—ranging from hundreds of thousands to potentially millions of code modifications—can create considerable performance bottlenecks. These challenges are especially pronounced when processing large repositories with extensive commit histories.

### 4.1.1 Processing Patterns

Two primary approaches to managing this large-scale data emerge:

1. **In-Memory Graph Construction:** One approach is to build and maintain a graph of all hunks in memory, where each node represents a hunk, and edges represent modifications to that hunk over time. This method has the advantage of providing fast access and update times, allowing for high-performance TTM calculations due to reduced I/O overhead. However, this approach introduces significant memory consumption, particularly as the repository size and history grow. Repositories with long commit histories and a high volume of changes may experience memory pressure, leading to inefficient use of system resources and potential failure due to memory exhaustion. As a result, this method is most feasible for repositories of moderate size or when sufficient computational resources (such as high-memory servers) are available.

2. **Local Data Store Approach:** An alternative approach is to store hunk data in a local database or file-based data store, where hunks can be indexed and referenced as needed. The advantage of this method is its scalability: by offloading data storage to disk, memory constraints are avoided. However, there are two key performance penalties associated with this approach.

    - First, even when using a local store, disk I/O introduces latency, particularly for write operations. Accessing data on disk is inherently slower than accessing data in memory, and this delay can become a limiting factor as the size of the repository grows.
    - Second, network-based remote data stores introduce even greater performance penalties due to the latency involved in network communication, making them suboptimal for this use case. As a result, a local data store is generally preferable, but the trade-off is slower access times relative to in-memory processing.

The size of the local data store itself becomes a factor. For repositories with a long history of commits or large numbers of contributors, the amount of data written to disk can grow substantially, leading to concerns about disk space availability and the additional overhead associated with managing a large data store.

### 4.1.2 Runtime and Efficiency Concerns

These processing challenges translate directly into concerns about the overall runtime required to calculate TTM across a large repository. Inefficient processing strategies, whether due to memory bottlenecks or slow disk I/O, can result in prohibitively long runtimes, rendering the durability analysis impractical for real-world use cases. Without optimizations in data access patterns, indexing strategies, or I/O operations, calculating TTM for every hunk in repositories of significant size could take hours or even days.

To mitigate these concerns, it may be necessary to explore hybrid approaches. For instance, a combination of in-memory processing for recent commits (where memory consumption is lower) and local data store usage for older parts of the repository may strike a balance between performance and scalability. Additionally, leveraging modern techniques such as lazy-loading data from the local store, caching frequently accessed hunks, and parallelizing I/O operations can help optimize the processing pipeline.

Ultimately, addressing these challenges is critical to ensuring the practical application of TTM in large-scale software systems. Advances in memory management, data storage strategies, and processing architectures will be essential to enable efficient durability analysis in modern software repositories.

### 4.1.3 Growth of Processing Time in Relation to Hunk Changes and Repository History

The time complexity of processing software contributions to calculate the Time to Modification (TTM) is inherently influenced by several factors, including the total number of hunk changes, the number of developers contributing to the repository, and the length of the repository's temporal history. As the size of the repository and the complexity of its commit history increase, the processing time for TTM calculations grows non-linearly due to the interplay between these variables.

*Mathematical Expression of Processing Time Growth*

Let the total number of hunk changes in the repository be denoted by $H$, the number of developers contributing to the repository by $D$, and the length of the repository's temporal history (measured by the total number of commits over time) by $T$. The processing time, $P(H, D, T)$, can be approximated by the following formula:

$$P(H, D, T) = \propto H \log(D) * \log(T)$$

Where:

- $H$ represents the total number of hunk changes that need to be processed.
- $log(D)$ reflects the logarithmic scaling effect of the number of developers. As the number of developers increases, the likelihood of overlapping changes in the same file or hunk increases, but this effect scales logarithmically due to the tendency of developers to work in parallel on different parts of the codebase.

- *log(T)* represents the effect of the repository's temporal history. With each additional commit over time, the system must traverse and compare an increasing number of historical states to calculate TTM, but this also scales logarithmically due to the way Git stores and manages deltas between commits.
- α is a constant of proportionality dependent on the specific system architecture, hardware, and data access speeds.

*Analysis of the Formula*

The formula highlights that while the total number of hunk changes *H* grows linearly with the size of the repository (i.e., as new commits and code contributions are made), the effect of both the number of developers *D* and the repository's temporal history *T* introduces logarithmic factors into the overall processing time.

- **Linear Growth with Hunk Changes:** The total number of hunks *H* contributes directly to the processing time. Each hunk must be processed to compute its TTM, and as the number of hunks increases, processing time grows linearly. This linear relationship can create significant overhead in large repositories with high commit activity or substantial code churn.
- **Logarithmic Growth with Developers:** As more developers contribute to the repository, the complexity of tracking the relationships between hunks increases. However, the logarithmic factor *log(D)* suggests that this complexity increases at a decreasing rate, owing to the fact that many developers work independently on separate portions of the code. Therefore, while adding developers introduces additional complexity, its impact on processing time is not exponential.
- **Logarithmic Growth with Temporal History:** Similarly, as the repository's temporal history T increases, each additional commit introduces a new state to consider in the TTM calculations. However, the efficiency of Git's delta storage model ensures that the number of comparisons grows logarithmically, rather than linearly, with the number of commits. This reflects the efficient traversal mechanisms used by Git when performing operations like blame and log analysis.

*Weaknesses of the Proposed Mathematical Expression*

While the formula $(H, D, T) = \propto H \log(D) * \log(T)$ provides a useful approximation, it simplifies several critical factors:

- **Developer Interaction:** The logarithmic scaling with developers ($\log(D)$) assumes parallel, independent work, but fails to account for increasing coordination overhead, merge conflicts, and code ownership issues, which may result in more complex behavior.
- **Repository Structure:** The formula overlooks the complexity introduced by branching, merging, and deeply entangled histories. Such structures can deviate from logarithmic behavior, especially when frequent merges or complex rebases occur.
- **Hunk Size Variability:** The linear scaling of hunks (*H*) assumes uniform processing time across hunks, but in reality, larger or more complex changes may introduce non-linear growth in processing time.
- **Temporal History:** While Git's delta compression is efficient, some operations (e.g., blame, rebase) may not scale logarithmically, particularly with deeply nested commit histories or complex merge scenarios.
- **System-Specific Factors:** The constant $α$ abstracts hardware and system dependencies. However, processing time may deviate in I/O-bound or memory-constrained environments, where cache performance and disk access play a larger role.

The formula provides a high-level estimate but may not fully capture real-world complexities in large, dynamic repositories.

*Implications for Large Repositories*

Given the formula, it is clear that for large repositories with extensive commit histories and many contributors, the processing time for calculating TTM can become a significant bottleneck. As H increases, the processing time grows proportionally, and while the logarithmic effects of D and T help mitigate exponential growth, they still contribute to increasing the overall runtime.

To address these challenges, optimizations in data access, parallel processing, and memory management are necessary. In particular, reducing the constant α through hardware improvements or more efficient algorithms can help manage processing time, especially in repositories where H, D, and T are all large.

This formal expression demonstrates that while the number of hunk changes exerts the most direct influence on processing time, both the number of developers and the length of the repository's history also affect processing efficiency. The logarithmic scaling effects of D and T suggest that repositories with large teams or long histories can still maintain reasonable processing times, provided that appropriate optimizations are in place. However, as repositories continue to grow in size and complexity, the importance of addressing performance challenges in TTM calculation becomes paramount.

## 4.2 Evaluating Developer Performance Using MTTM

The use of Mean Time to Modification (MTTM) metrics offers a powerful and objective method for evaluating the performance and quality of software contributions in a production setting. By measuring how long code written by individual developers remains unmodified, managers can gain insights into both the durability of code and the effectiveness of the development team. This approach can be leveraged to make informed decisions about software quality, developer performance, and project management.

### 4.2.1 MTTM as a Measure of Developer Contribution Quality

The MTTM reflects how long, on average, a developer's code remains unchanged before requiring modification. In general, higher MTTM values suggest that a developer's contributions are durable, which may indicate that their code is of higher quality, requiring fewer immediate fixes or revisions. Conversely, lower MTTM values could signal that code is being modified soon after introduction, potentially indicating that the code requires improvements, is prone to defects, or is less aligned with the overall architecture of the system.

In a production environment, this metric can be used in various ways:

- **Assessing Code Stability:** Developers with consistently high MTTM values likely produce code that is stable and resilient to changes. This can be an indicator of well-considered designs and thoughtful implementation. Managers can use these insights to identify key contributors whose work has a lasting positive impact on the codebase.
- **Identifying Areas for Improvement:** Lower MTTM values might point to developers who are introducing code that frequently requires modification. This could indicate issues with the developer's approach to problem-solving, the need for better requirements clarity, or an opportunity for mentoring and skill development. Regular review of these metrics can help identify areas where further training or support might be needed.
- **Comparing Team Performance:** MTTM can be used to compare performance across teams or individual contributors. Averages can be calculated for different project teams, and trends over time can be analyzed to assess how code stability is evolving. Teams with significantly lower MTTM scores may benefit from process improvements, code reviews, or refactoring initiatives.

### 4.2.2 Impact on Software Engineering Management

The integration of MTTM metrics into routine performance reviews and project management processes offers several practical benefits:

1. Objective Performance Evaluation: Traditional performance evaluations in software engineering can sometimes be subjective, relying heavily on peer feedback or code review quality. MTTM provides a quantifiable, data-driven measure of performance, allowing managers to assess developers based on concrete metrics that reflect their actual contributions to the codebase.

    - Data-Driven Decisions: MTTM helps reduce bias by focusing on measurable outcomes, ensuring that high-performing developers are recognized for producing durable and maintainable code. This enables more fair and objective evaluations across the team.

2. Encouraging Long-Term Thinking: Knowing that their contributions are being measured for durability may encourage developers to take a more thoughtful approach to writing code. This could shift focus from quick fixes or short-term solutions toward more sustainable, long-term implementations that are less prone to frequent modification. By emphasizing MTTM, managers can foster a culture that values the longevity and maintainability of the code, leading to more stable software systems.

3. Improving Project Efficiency: Software engineering teams are often evaluated based on delivery speed and the number of features implemented. However, frequent modifications to code after initial deployment can create technical debt, increasing the cost of future changes. By integrating MTTM into the performance evaluation process, teams can balance speed with quality, ensuring that code is more resilient and reduces the need for future rework.

    - Fewer Reworks, Higher Efficiency: As developers produce higher-quality, longer-lasting code, the team spends less time revisiting and revising previous work. This reduction in rework can significantly improve project efficiency, freeing up time for new features or innovation.

4. Monitoring Technical Debt: MTTM metrics can help managers keep track of areas in the codebase where technical debt is accumulating. Code that is frequently modified soon after its introduction may indicate an area of the system that is poorly designed or difficult to maintain. By monitoring MTTM trends, managers can proactively identify and address technical debt before it becomes unmanageable, leading to better overall project health.

5. Enhancing Developer Motivation: MTTM offers an additional avenue for recognizing developers who consistently deliver durable, high-quality code. This can be a powerful motivator, as it provides developers with a clear and measurable way to demonstrate their impact on the project. High-performing developers can be acknowledged not just for the quantity of their contributions but also for the long-term stability of their work, leading to greater job satisfaction and career development.

6. Encouraging More Careful Team and Paired Reviews: The use of MTTM metrics may indirectly encourage teams to adopt more careful code review practices, particularly in paired or collaborative settings. When developers know that their contributions will be measured not only for immediate correctness but also for long-term stability, they are more likely to engage in thorough, detailed reviews. Pair programming and team reviews, in this context, become crucial tools for ensuring code longevity.

- Promoting Accountability: Paired reviews foster a sense of shared responsibility among team members, encouraging them to scrutinize each other's work with an eye toward long-term maintainability. The awareness that code changes will be evaluated by MTTM metrics incentivizes developers to avoid rushing through reviews and promotes a culture of accountability.

- Reducing Oversights: Careful team reviews reduce the likelihood of oversights that could lead to short-lived or unstable code. With MTTM metrics highlighting code that requires frequent modifications, the pressure to ensure that code is solid from the outset increases, leading to more rigorous peer reviews and collaborative discussions around design and implementation choices.

- Fostering a Culture of Quality: By integrating MTTM into review processes, teams can create a culture that prioritizes quality over quantity. Developers and reviewers alike become more attuned to identifying potential issues that might lead to frequent modifications down the line, such as poor design decisions, unclear logic, or insufficient testing. As a result, reviews become a mechanism not just for catching immediate bugs, but for improving the overall durability of the codebase.

Incorporating MTTM metrics into the evaluation of developer performance and project management can provide actionable insights into the quality and stability of code contributions. By focusing on the durability of code, managers can better assess the long-term value that developers bring to a project, while also identifying opportunities for improvement in coding practices, team dynamics, and process efficiency. When used thoughtfully, MTTM metrics can enhance both individual performance management and the overall quality of the software engineering process.

***Practical Considerations and Limitations***

While MTTM is a valuable tool, it should not be used in isolation to assess developer performance. There are several practical considerations to keep in mind:

- **Contextual Factors:** MTTM must be interpreted in the context of the specific project or feature being worked on. Some projects, particularly those involving experimental or exploratory code, may naturally have lower MTTM values, as frequent iterations and modifications are expected. In such cases, lower MTTM does not necessarily indicate poor performance.
- **Collaboration vs. Individual Performance:** Software development is often a collaborative process, and code modifications may occur due to team-wide decisions rather than individual developer errors. Managers should ensure that MTTM metrics are evaluated alongside other performance indicators, such as teamwork, adherence to architectural standards, and the complexity of the tasks involved.
- **Outlier Management:** Extremely high or low MTTM values may represent outliers and should be treated with caution. For example, code that remains unmodified for a very long time could be a sign of code that has been abandoned or neglected rather than an indication of high quality. Similarly, very low MTTM values may result from changes in external dependencies or shifting project requirements rather than poor development practices.

Incorporating MTTM metrics into the evaluation of developer performance and project management can provide actionable insights into the quality and stability of code contributions. By focusing on the durability of code, managers can better assess the long-term value that developers bring to a project, while also identifying opportunities for improvement in coding practices, team dynamics, and process efficiency. When used thoughtfully, MTTM metrics can enhance both individual performance management and the overall quality of the software engineering process.

# 5 Future Work

Time to Modification (TTM) Theory provides a robust foundation for understanding code durability and developer performance. However, the application of this theory can be expanded in several key areas to improve its utility and accuracy. Drawing from existing research, including Contribution Rate Imputation Theory (CRIM) and the Time-Delta Method (TDM) by Bishop et al. [23] [1], future work can leverage these approaches to enhance the theoretical and practical framework of TTM. This section outlines key directions for future research, focusing on integrating TTM with additional metrics, expanding the scope of analysis, and applying predictive modeling techniques.

## 5.1 Integrating TTM with Contribution Rate Imputation Theory

Contribution Rate Imputation Theory (CRIM), as proposed by Bishop et al., estimates developer effort by analyzing historical development data and calculating unobserved work periods using metrics such as cyclomatic complexity and Levenshtein distance [23]. Integrating TTM with CRIM offers a promising approach to refining the measurement of code durability. CRIM's emphasis on imputation methods for estimating developer effort could be extended to estimate the expected lifespan of code contributions, offering deeper insights into both the quality and effort involved in a developer's contributions.

- **Imputing Contribution Durability:** By incorporating TTM into CRIM's imputation framework, we can estimate the expected Time to Modification based on historical patterns of contribution rates. This would allow for the prediction of code durability in cases where direct measurement of TTM is not feasible (e.g., newly introduced code). Imputing TTM would be especially useful in dynamic environments, enabling better resource planning and project forecasting.
- **Comparing Contribution and Durability Rates:** The combination of TTM and CRIM also opens new avenues for comparing contribution rates with code durability. For instance, contributions with higher imputed rates of effort might correlate with longer TTM values, indicating more durable code. Conversely, low-effort contributions may lead to shorter TTM, reflecting code that requires frequent modifications. This correlation would provide a comprehensive picture of how contribution rates influence long-term code stability.

## 5.2 Expanding the Time-Delta Method to Account for Multiple Modifications

The Time-Delta Method (TDM) [1] offers a method for estimating developer effort by analyzing commit intervals and associating them with code metrics. The TDM framework can be extended to account for multiple modifications to the same code segment, thereby enhancing the TTM model's ability to capture long-term contribution durability.

- **Recurrent Modifications:** Currently, TTM focuses on the first modification of a hunk. However, the same code may be modified multiple times, and capturing these subsequent modifications is critical for understanding long-term durability trends. Future research could build on TDM's ability to track intervals between commits and modifications to construct a more comprehensive model of how frequently code is changed over time.
- **Time-Weighted Durability:** A potential extension of TDM and TTM involves calculating time-weighted durability, where the durability score of a hunk is adjusted based on the timing and frequency of modifications. Code that is modified multiple times in quick succession may exhibit lower durability than code modified infrequently, even if both have the same initial TTM. This approach would allow for a more nuanced understanding of code stability beyond the first modification.

## 5.3 Developing Predictive Models for TTM

Leveraging machine learning and predictive modeling could significantly enhance the utility of TTM, allowing for real-time prediction of code durability based on a wide range of factors.

- **Feature Engineering:** To develop accurate predictive models, future work should focus on identifying key features that influence TTM. These features could include commit size, code complexity (e.g., cyclomatic complexity), developer experience, and contribution history. By building a robust feature set, machine learning models can predict the likely TTM of new code contributions based on historical data.
- **Integration with CI/CD Pipelines:** Predictive models for TTM could be integrated into continuous integration/continuous deployment (CI/CD) pipelines. By predicting the durability of new code before it is merged into the main branch, teams can proactively identify potential areas of instability and trigger additional quality checks or refactoring efforts. This integration would help improve code stability in production environments.

### 5.4 Combining TTM with Software Quality Metrics

Combining TTM with other established software quality metrics offers an opportunity to build a more comprehensive framework for evaluating code quality and developer performance. Cyclomatic complexity, Levenshtein distance, and Contribution Rate Imputation can all be used to enhance the interpretation of TTM.

- **Multivariate Analysis:** Using a multivariate approach, TTM can be analyzed alongside other metrics such as cyclomatic complexity and Levenshtein distance (as described in CRIM). This combination would allow for a richer analysis of code durability, helping to identify whether complex or highly modified code exhibits shorter or longer TTM values. For example, high cyclomatic complexity combined with low TTM may indicate code that is both complex and frequently unstable.
- **Holistic Quality Models:** By combining TTM with Contribution Rate Imputation and the Time-Delta Method, a holistic model of code quality can be developed. Such a model would take into account both the time spent on contributions and the durability of the resulting code. This could lead to more informed decision-making in software development projects, particularly regarding resource allocation and developer performance evaluation.

### 5.5 Benchmarking TTM Across Different Project Types

As TTM metrics are applied in production settings, establishing benchmarks for different types of software projects will become increasingly important. These benchmarks could provide valuable context for interpreting TTM values across various industries and project types.

- **Industry-Specific Benchmarks:** Future research could focus on creating benchmarks for TTM in different industries (e.g., finance, healthcare, gaming) to account for varying development practices and stability expectations. By understanding how TTM varies across sectors, teams can more accurately assess the durability of their code relative to industry standards.
- **Project Lifecycle Benchmarks:** Establishing benchmarks for TTM across different phases of the software development lifecycle (e.g., feature development, maintenance, bug fixing) would allow teams to interpret TTM in a way that aligns with the goals of each project phase.

### 5.6 Benchmarking TTM Against Cybersecurity Metrics

In addition to benchmarking TTM across project types, it is important to benchmark it against cybersecurity metrics, such as mean time to detect (MTTD) vulnerabilities and mean time to remediate (MTTR). By comparing TTM with MTTD, teams can assess whether code that remains unmodified for long periods is more likely to harbor undetected vulnerabilities. Additionally, correlating TTM with MTTR can reveal how quickly stable code can be patched once a vulnerability is identified.

- Stable code with high TTM values may require more regular security audits to ensure vulnerabilities are detected and patched promptly. Tools like SonarQube can be integrated with TTM metrics to flag segments that may require more scrutiny from a security perspective. Benchmarking TTM against cybersecurity metrics ensures that durable code is not only maintainable but also secure, especially in industries with high cybersecurity requirements like finance and healthcare.

TTM Theory has established a strong foundation for measuring code durability, but significant opportunities remain for future research. By integrating TTM with the Contribution Rate Imputation Method [23], expanding the Time-Delta Method [1], developing predictive models, and exploring the role of cross-functional teams, we can continue to refine this theory and make it a valuable tool for software engineering management and research. Through these expansions, TTM can evolve into a comprehensive framework for understanding and improving the quality and durability of software contributions in diverse development environments.

## 6 Conclusions

The Time to Modification (TTM) Theory presents a novel approach to assessing the quality and durability of software contributions by focusing on the temporal stability of code. TTM, as a measurable indicator, offers valuable insights into the longevity of a code segment before its first modification, serving as a proxy for contribution quality. This research demonstrates that longer TTM values generally correlate with more stable, well-designed, and maintainable code, providing software engineers with a data-driven mechanism to evaluate developer performance and code sustainability.

By introducing TTM into the software engineering domain, this theory addresses a key gap in traditional code quality metrics, which often overlook temporal aspects. The ability to track how long code remains unmodified gives development teams a clearer picture of the long-term viability of their contributions. Unlike static code analysis or code complexity metrics, TTM focuses on the evolution of the codebase over time, offering a dynamic perspective on software maintenance and highlighting areas where code may require refactoring or closer attention.

The broader implications of TTM Theory for the software industry are substantial. As software systems become increasingly complex, managing technical debt and ensuring long-term maintainability are critical. TTM offers an actionable metric to predict and mitigate potential maintenance issues early in the development lifecycle, reducing long-term costs. This theory can be integrated into continuous integration/continuous deployment (CI/CD) pipelines to monitor the stability of code contributions in real-time, helping teams to identify unstable code segments before they become costly to maintain.

Additionally, TTM Theory provides a robust framework for comparing individual developer contributions and team performance. With the ability to measure the temporal durability of code, managers can make more informed decisions about resource allocation, developer training, and project planning. It also encourages developers to prioritize maintainability and long-term thinking in their contributions, as they are aware that their work will be evaluated not just on immediate functionality, but on its ability to endure over time without frequent modifications.

In practical terms, TTM's integration with other software quality metrics, such as cyclomatic complexity and code churn, allows for a more comprehensive assessment of code quality. By combining these measures, teams can gain deeper insights into the relationships between code complexity, frequency of modifications, and overall durability. Furthermore, TTM's application in machine learning models opens the possibility for predictive analysis, enabling teams to forecast the stability of new code based on historical data.

Despite the clear benefits, the computational complexity of calculating TTM at scale presents challenges, particularly for large repositories with extensive commit histories. Optimizing data access patterns and improving processing architectures will be critical to ensuring the practical application of this theory in real-world, large-scale software systems. Future work will need to explore more efficient ways to manage this data, particularly as repositories grow in size and complexity.

TTM Theory offers a valuable new lens through which software quality can be evaluated. Its focus on temporal durability provides insights that complement traditional metrics, enhancing both developer performance assessments and the maintainability of software systems. The application of TTM in modern software practices has the potential to significantly improve long-term software quality, reduce maintenance costs, and contribute to more sustainable software engineering practices.

## Acknowledgements

The authors would like to express their deepest gratitude to the Systems Engineering Department at Colorado State University, Fort Collins, for their invaluable support in making this research possible.

*During the preparation of this work, artificial intelligence was used in the research and writing processes. After using these services, the content was reviewed and edited as needed and full responsibility is taken for the content of this work.*

# References

[1] V. Bishop Iii, "Time-Delta Method for Measuring Software Development Contribution Rates," Mountain Scholar : Digital Collections of Colorado, 2024. [Online]. Available: https://hdl.handle.net/10217/239202

[2] R. Plösch *et al.*, "The EMISQ Method and Its Tool Support-Expert-Based Evaluation of Internal Software Quality," *Innovations in Systems and Software Engineering,* vol. 4, no. 1, pp. 3-15, 2008, doi: 10.1007/s11334-007-0039-7.

[3] I. I. Hadyan, D. S. Kusumo, and J. H. Husen, "Implementing Continuous Code Quality for Code Quality Development in the Scrum Team," *Jipi (Jurnal Ilmiah Penelitian Dan Pembelajaran Informatika),* vol. 7, no. 3, pp. 728-734, 2022, doi: 10.29100/jipi.v7i3.3066.

[4] K. K. Zaw, H. W. Hnin, K. Y. Kyaw, and N. Funabiki, "Software Quality Metrics Calculations for Java Programming Learning Assistant System," 2020, doi: 10.1109/icca49400.2020.9022823.

[5] C. Kothapalli *et al.*, "Continual Monitoring of Code Quality," 2011, doi: 10.1145/1953355.1953379.

[6] E. Jharko, "Evaluation of the Quality of a Program Code for High Operation Risk Plants," *Ifac Proceedings Volumes,* vol. 47, no. 3, pp. 8060-8065, 2014, doi: 10.3182/20140824-6-za-1003.02140.

[7] D. Kapse, K. Kabra, M. Chopade, and K. Palhal, "Measuring Code Quality," *Ijarcce,* pp. 423-425, 2015, doi: 10.17148/ijarcce.2015.43101.

[8] M. Foucault, C. Teyton, D. Lo, X. Blanc, and J.-R. Falleri, "On the Usefulness of Ownership Metrics in Open-Source Software Projects," *Information and Software Technology,* vol. 64, pp. 102-112, 2015, doi: 10.1016/j.infsof.2015.01.013.

[9] S. Butler, M. Wermelinger, Y. Yu, and H. Sharp, "Exploring the Influence of Identifier Names on Code Quality: An Empirical Study," 2010, doi: 10.1109/csmr.2010.27.

[10] T. Guo, H. Bai, Y. Gong, Y. Wang, and D. Jin, "An Experimental Study on Attribute Validity of Code Quality Evaluation Model," *Tehnicki Vjesnik - Technical Gazette,* vol. 29, no. 2, 2022, doi: 10.17559/tv-20211029115834.

[11] N. Nagappan and T. Ball, "Use of Relative Code Churn Measures to Predict System Defect Density," p. 284, 2005, doi: 10.1145/1062455.1062514.

[12] A. Meneely and O. Williams, "Interactive Churn Metrics," *Acm Sigsoft Software Engineering Notes,* vol. 37, no. 6, pp. 1-6, 2012, doi: 10.1145/2382756.2382785.

[13] C. Hota, L. Kumar, and L. B. M. Neti, "An Empirical Analysis on Effectiveness of Source Code Metrics for Aging Related Bug Prediction," 2019, doi: 10.18293/dmsviva2019-022.

[14] C. Chen and M. Shoga, "A Large Empirical Study on Automatically Classifying Software Maintainability Concerns From Issue Summaries," *Advances in Science Technology and Engineering Systems Journal,* vol. 6, no. 2, pp. 161-174, 2021, doi: 10.25046/aj060219.

[15] D. W. Keenan, D. Greer, and D. Cutting, "An Investigation Of Entropy And Refactoring In Software Evolution," pp. 282-297, 2022, doi: 10.1007/978-3-031-21388-5_20.

[16] G. Szoke, C. Nagy, R. Ferenc, and T. Gyimóthy, "A Case Study of Refactoring Large-Scale Industrial Systems to Efficiently Improve Source Code Quality," pp. 524-540, 2014, doi: 10.1007/978-3-319-09156-3_37.

[17] K. D. Welker, P. Oman, and G. G. Atkinson, "Development and Application of an Automated Source Code Maintainability Index," *Journal of Software Maintenance*

*Research and Practice,* vol. 9, no. 3, pp. 127-159, 1997, doi: 10.1002/(sici)1096-908x(199705)9:3<127::aid-smr149>3.0.co;2-s.

[18]  N. Haderer, F. Khomh, and G. Antoniol, "SQUANER: A Framework for Monitoring the Quality of Software Systems," 2010, doi: 10.1109/icsm.2010.5609684.

[19]  A. Thahseen, "Analyzing the Impact of Software Testing on Software Maintainability," 2023, doi: 10.21203/rs.3.rs-2927364/v1.

[20]  L. Vidács, F. Horváth, D. Tengeri, and A. Beszedes, "Assessing the Test Suite of a Large System Based on Code Coverage, Efficiency and Uniqueness," 2016, doi: 10.1109/saner.2016.69.

[21]  S. Saklani, "Software Quality Prediction Using Machine Learning Techniques and Source Code Metrics: A Review," *International Journal of Advanced Research in Computer Science,* vol. 13, no. 06, pp. 12-25, 2022, doi: 10.26483/ijarcs.v13i6.6918.

[22]  Z. M. Jiang and A. E. Hassan, "Examining the Evolution of Code Comments in PostgreSQL," 2006, doi: 10.1145/1137983.1138030.

[23]  V. Bishop, III and S. Simske, "Contribution Rate Imputation Theory: A Conceptual Model," *arXiv,* vol. 2410.09285, p. 14, 2024, doi: 10.48550/arXiv.2410.09285.